\documentclass[12pt,a4paper]{article}
\usepackage{amssymb}
\usepackage{amsfonts}
\usepackage{amsmath}
\usepackage{color}
\usepackage{graphicx}
\usepackage{hyperref}
\begin{document}

\renewcommand{\vec}[1]{\boldsymbol{#1}}
\newcommand{\ATfrac}[2]{#1\left/#2\right.}
\newcommand{\KEO}[1]{T^{}_{#1}}
\date{\today}

\title{Classification Scheme for Kinetic Energy Operators with Position-Dependent Mass}

\author{A. Trabelsi$^\star$,
 F. Madouri,
 A. Merdaci \\
 and A. Almatar}
\maketitle 

\begin{center}
{\small \textsl{Physics Department, College of Science, King Faisal University\\ P.O. 380 Ahsaa 31982, Saudi Arabia}}
\end{center}

\small{$^\star$Email: atrabelsi@kfu.edu.sa}

\begin{abstract}
In this paper we present a complete classification scheme for kinetic energy operators (KEO's) describing a particle endowed with  position-dependent mass (PDM). We first present a generalized formulation of KEO's with PDM and show that it is equivalent to three-parameter linear formulation. This reduces further to two independent parameters under Hermiticity condition. Based on this linear formulation we prove that, contrary to what was widely believed, von Roos family is not the most general ordering. We found an entire new family of Hermitian KEO's that does not fit into von Roos ordering.  We were able to construct all the Hermitian KEO's and classify them into two-parameter classes.   As one application, we solve the puzzling case of Yang and Yee KEO. We also find, under certain conditions, some kind of duality between von Roos ordering and one of the new classes.
\end{abstract}



\section*{Introduction}

The Schrodinger equation subject to PDM was for decades an intensive field of fundamental research activity as demonstrated by already  published works on the subject.  In addition of being a descriptive model for many physical phenomena of current interest such as electronic properties of non-uniform semiconductors and heterojunctions \cite{geller1993quantum,ring2005nuclear,%
ganguly2006study,plastino2001bohmian,%
saavedra1994effective,puente1994dipole,bendaniel1966space,%
li1993band,bastard1981superlattice,gora1969theory,%
zhu1983interface,morrow1984model}, the Schrodinger equation with PDM is, by itself, a fundamental problem which is far from being completely understood.
It raises some important conceptual problems, such as the Galilean invariance of the theory~\cite{levy1995position,von1983position}, the boundary and continuity conditions at abrupt interfaces~\cite{levy1995position,morrow1984model,zhu1983interface}, calculation of Green's function~\cite{alhaidari2003nonrelativistic, chetouani1995green}, implementation of path-integral technique~\cite{chetouani1995green,yung1994derivation}, potential algebra technique~\cite{roy2005effective,jana2009potential}, supersymmetric formalism~\cite{quesne2006first,milanovic1999generation,borges1988quantization,%
plastino1999supersymmetric,gonul2002supersymmetric} and shape invariance technique~\cite{plastino1999supersymmetric,ganguly2007shape,%
gonul2002supersymmetric,quesne2009point}, and of course exact solvability~\cite{dekar1998exactly,%
plastino2000bound,alhaidari2002solutions,%
yu2004exactly,mustafa2009spherical,%
bagchi2005deformed,dutra2000exact,bagchi2004general}.

However, by far the most challenging problem still to be resolved is the ambiguity associated with momentum- and mass-operators non-commutativity  that is at the origin of all the preceding issues~\cite{von1983position,morrow1984model,%
levy1995position,%
zhu1983interface,bastard1981superlattice,gora1969theory,%
dutra2000exact,bagchi2004general,li1993band,%
thomsen1989operator,cavalcante1997form}. Indeed, when the mass $m(\vec{r})$ is position-dependent, it does no longer commute with the momentum operator~$\vec{p}$. The standard nonrelativistic KEO of quantum mechanics
\begin{align}\label{Tstd}
T=\frac{1}{2}  \frac{\vec{p}^2}{m_0}~,
\end{align}
valid for a constant mass $m_0$, becomes ill-defined for PDM. We have to determine how to order the mass relative to the momentum operators in order to generalize the usual form~(\ref{Tstd}). This is a particular, nonetheless important, case of the long standing ordering ambiguity problem in quantum mechanics, see for instance
the excellent review by Shewell~\cite{shewell1959formation}. This ordering ambiguity is not just a conceptual problem but it appears to be relevant to many physical systems endowed with PDM~\cite{von1983position,levy1995position,levy1992elementary,%
morrow1984model,%
galbraith1988envelope,brezini1995model}. The Hermiticity condition of the Hamiltonian, current-density conservation, comparison between experiments and theoretical results were unable to indicate conclusively a unique form for the KEO with a position-dependent effective mass. The question of the exact form of the KEO, if there exists one, is hitherto an open issue which has advanced very few along the last decades.

The rest of paper is organized as follows.  We start by reviewing the existing KEO's and some important concepts, such as ordering ambiguity parameters. We mathematically formulate the problem within a very general and model independent framework. Then we derive a complete classification scheme for KEO's with PDM. We prove that, in addition to the well known  von Roos family there is an other new family of Hermitian  KEO's with PDM. We proceed to the full characterization of this family and classify  its KEO's into two-parameter classes.  Then we investigate the duality between the two families.  Finally, in the last section the
main results are summarized and some conclusions are drawn.

\section{Ordering Ambiguity Problem}

To take into account the position dependence of the effective mass, various expressions of KEO's were suggested in the literature and arguments were put forward to support them. In view of the great diversity of proposed KEO's and the scope o this work,  an as complete as possible review of earlier works is indicated. This review is meant to highlight the need for a model independent classification scheme to handel those KEO's.

BenDaniel and Duke~\cite{bendaniel1966space} were the first to examine  the problem of KEO with PDM in the effective-mass approximation. They have concluded that in order to ensure current conservation, it is necessary to replace the ordinary KEO in~(\ref{Tstd})  by
\begin{align}\label{eq:TBDD}
T_{\rm BDD}=\frac{1}{2}  \vec{p}
{\frac{1}{m}} \vec{p}~.
\end{align}
This was apparently the  first Hermitian KEO  with PDM.  It is the most used for analytical calculations and for which some physical evidences have been provided, although recent works have suggested that other operators are more suitable. Later on, Gora and Williams~\cite{gora1969theory} proposed  another pertinent KEO to model a binary alloy of a position-dependent composition. They have used Slater's method and derived a two-term KEO
\begin{align}\label{eq:TGW}
T_{\rm GW}=\frac{1}{4}\left[ {\frac{1}{m}} \vec{p}^2
  + \vec{p}^2{\frac{1}{m}} \right]~.
\end{align}
Expression (\ref{eq:TGW}), also adopted by Bastard \emph{et al.}~\cite{bastard1975landau}, had pointed out early on the problem of non-uniqueness of KEO with PDM.

Another major ansatz for maintaining Hermiticity of the KEO was suggested  by  Zhu and Kroemer~\cite{zhu1983interface}. In order to reformulate the connection rule problem on the two sides of an heterojunction, they have proposed to partition the mass and order the KEO in the follwing way
\begin{align}\label{eq:ZK}
T_{\rm ZK}=\frac{1}{2}\,{\frac{1}{\sqrt{m}}}\,\vec{p}^2
{\frac{1}{\sqrt{m}}}~.
\end{align}
The same form was derived later by Cavalcante \emph{et al.}~\cite{cavalcante1997form} through the nonrelativistic limit of the Dirac Hamiltonian with PDM by means of a Foldy-Wouthuysen transformation.

In an attempt to cope with the non-uniqueness representation of the KEO, von~Roos~\cite{von1983position} proposed a two-parameter family of KEO's which has a built-in Hermiticity. It contains the above one- and two-term alternative forms  as special cases. It is expressed as follows
\begin{align}\label{eq:TVR}
T_{\rm vR}=\frac{1}{4}\left[ m^{\alpha} \vec{p}\,m^{\beta } \vec{p}\,m^{\gamma} +m^{\gamma} \vec{p }\,m^{\beta} \vec{p}\,m^{\alpha} \right]~,
\end{align}
where $\alpha$, $\beta$ and $\gamma$, called von~Roos ordering ambiguity parameters, are real and  constrained by the condition
\begin{align}\label{abg1}
\alpha+\beta+\gamma=-1~,
\end{align}
to match the classical kinetic energy expression. Though, only two independent ordering ambiguity parameters are actually needed. Obviously, changing the values of $\alpha$, $\beta$ and $\gamma$ would change the resulting expression of $T_{\rm vR}$. For consistency, it is also supposed, but not mathematically required,  that
\begin{align}\label{0abg1}
0\geq\alpha,\beta,\gamma\geq-1~.
\end{align}
Unfortunately there is no universally agreed  single set of values for ordering ambiguity parameters that emerges from this formulation. For a given choice of ordering ambiguity parameters $\{\alpha,\beta,\gamma\}$ we may have up to three different orderings. For instance, the set $\{0,0,-1\}$ leads to BDD ordering  ($\alpha=0$, $\beta =-1$, $\gamma=0$) or  GW ordering   ($\alpha=0$, $\beta =0$, $ \gamma=-1$).

Morrow and Brownstein~\cite{morrow1984model},  addressing abrupt heterojunctions between two crystals with discontinuous step-like distribution of effective masse, have proposed a one-parameter subclass of von Roos ordering
\begin{align}\label{eq:TMB}
T_{\rm MB}=\frac{1}{2}  m^{\alpha} \vec{p}\,
m^{\beta} \vec{p} \,m^{\alpha}~,
\end{align}
with $2\alpha+\beta=-1$. Notice that $T_{\rm BDD}$ and $T_{\rm ZK}$ are respective special cases for $ \alpha=0$ and $ \alpha=-\frac12$. More recently  Mustafa and Mazharimousavi (MM) using a PDM-pseudo-momentum operator, have suggested another particular case with $ \alpha=-\frac14$~\cite{mustafa2007ordering}.
On the other hand, Dutra and Almeida~\cite{dutra2000exact} have proposed another one-parameter subclass
\begin{align}\label{eq:TLKDA}
T_{\rm LKDA}=\frac{1}{4}\left[ m^{\alpha} \vec{p}\,%
m^{\beta} \vec{p} + \vec{p}\,m^{\beta} \vec{p}\,m^{\alpha} \right]~,
\end{align}
by extending the Li and Kuhn ordering (LK) originally defined for $ \alpha=-\frac12$~\cite{li1993band}.
Furthermore, they have proved that LK-ordering is equivalent to the three-term  Weyl-ordering
\begin{align}\label{eq:Weyl}
T_{\rm W}=\frac{1}{8}\left[  {\frac{1}{m}}\vec{p}^{2} +2\,\vec{p}\,{\frac{1}{m}}\,\vec{p}+
\vec{p}^2{\frac{1}{m}}\right]~.
\end{align}
This ordering has been also derived using quantization in the phase-space path-integral framework~\cite{borges1988quantization}. An interesting feature of the LKDA subclass is that the following four-term KEO
\begin{align}\label{eq:TDA}
T_{\rm DA}=\frac{1}{4(\alpha +1)}\Bigg(& \alpha \left[ {\frac{1}{m}}\,\vec{p}^{2}+\vec{p}^{2}\,{\frac{1}{m}}\right]
+m^{\alpha }\vec{p}m^{\beta }\,\vec{p}
+\vec{p}\,m^{\beta }\,\vec{p}m^{\alpha }\Bigg)
\end{align}
dose not depend on the ordering ambiguity parameters as noticed in~\cite{dutra2000exact}. More recently Lima \textit{et al.}~\cite{lima2012yet} used  yet another three-term KEO
\begin{align}\label{eq:Lal}
T_{{\rm L}al.}=\frac{1}{6}\left[  {\frac{1}{m}}\vec{p}^{2} +\vec{p}{\frac{1}{m}}\vec{p}+
\vec{p}^2{\frac{1}{m}}\right]
\end{align}
to calculate the band structure for a quantum particle with mass varying periodically. This KEO which appears not to fit the two-term von Roos form, was found out to produce different results from BDD.

The above overview assesses the controversial problem of KEO ordering ambiguity and the inherent variation that this induces in both the formalism and the results.  There is clearly  a lack of a global formulation and consequently a complete classification of KEO's with PDM. One important question is at which extent we can continue adding more and more terms to the KEO.  We shall answer this in next section.

\section{General formalism}

In order to develop a general classification scheme for the already proposed KEO's and also other possible forms it would be useful to  work in a model-independent way. Many authors~\cite{levy1995position,thomsen1989operator,dutra2003remarks,%
plastino2001bohmian,gonul2002supersymmetric,bagchi2005deformed,bagchi2004general,%
thomsen1989operator,mustafa2007ordering,dutra2000exact,lima2012yet} have already used linear parametrization to represent von Roos KEO's. It was much more for technical ansatz rather than a fundamental approach.  We shall extend this approach to the most general form  of KEO with PDM.

The building block of KEO's with PDM has the general form $m^{\alpha}\vec{p}\,m^{\beta }\vec{p}\, m^{\gamma}$. More complicated building blocks as suggested for instance by~\cite{morrow1987effective} are certainly possible but there seems little to be gained in introducing more parameters than necessary at this stage. Allowing for an arbitrary number of building blocks, the general KEO would read
\begin{align}\label{TGEN}
T_{\vec{\alpha},\vec{\beta},\vec{\gamma}}= &\frac{1}{2}\sum_{i=1}^N w_i\: m^{\alpha_i}\vec{p}\,m^{\beta_i}\vec{p}\,m^{\gamma_i}~,
\end{align}
where $N$ is an arbitrary positive integer. To match the classical limit, the constant parameters $\vec{\alpha}=(\alpha_i)$, $\vec{\beta}=(\beta_i)$ and $\vec{\gamma}=(\gamma_i)$ should fulfill the constraint
\begin{align}\label{Constraint}
\vec{\alpha}+\vec{\beta}+\vec{\gamma}=(-1,\ldots,-1)~,
\end{align}
and the real weights $\vec{w}=(w_i)$ should sum to unity.
Analytical calculation shows that the form~(\ref{TGEN}) greatly  simplifies to
\begin{align}\label{TabgNH}
T_{\vec{\alpha},\vec{\beta},\vec{\gamma}}=   &\frac{1}{2}\vec{p}{\frac{1}{m}}\vec{p}
+(\overline{\gamma}-\overline{\alpha})\frac{i\hbar}{2}\vec{\nabla}{\frac{1}{m}}\cdot\vec{p} +\frac{\hbar^2}{2}\left[\overline{\gamma}\:\vec{\nabla }%
^{2}{\frac{1}{m}} +\overline{\alpha\gamma} \ATfrac{\left(\vec{\nabla}\frac{1}{m}\right)^{2}\!\!}{\frac{1}{m}}\right],
\end{align}
$\overline{\!X}$ denotes the $w_i$-weighted mean values:
\begin{align}\label{MeansAGAG}
 \overline{\!X}=\sum_{i=1}^N w_i X_i~.
\end{align}
The last term in (\ref{TabgNH}) acts as an effective  mass- and  ordering ambiguity parameter-dependent additional contribution to the original physical potential  originating from the momentum and mass-operator non-commutativity.

The KEO (\ref{TabgNH}) is not Hermitian because of the term containing  $\vec{\nabla}(1/m)\cdot\vec{p}$ operator. The Hermiticity condition requires this term to vanish, that is
\begin{align}\label{HermAG}
\overline{\alpha}=\overline{\gamma}~,
\end{align}
which we shall assume  henceforth. Notice that this is a much more general requirement than the one-to-one matching of $\alpha_i$ and $\gamma_j$ as suggested by von Roos and its generalization as proposed in~\cite{thomsen1989operator}.

Thus the $(3N-1)$-parameter dependent KEO in (\ref{TGEN}) reduces to a manifestly two-parameter dependent one, which we can rewrite  as
\begin{align}\label{Txizeta}
T_{\xi,\zeta}= &\frac{1}{2}\vec{p}{\frac{1}{m}}\vec{p}
+\frac{\hbar^{2}}{2}\left[ \xi  \vec{\nabla}^{2}\frac{1}{m} +\zeta \ATfrac{\left(\vec{\nabla}\frac{1}{m}\right)^{2}\!\!}{\frac{1}{m}}\right],
\end{align}
using the two linear ordering ambiguity  parameters
\begin{align}\label{xizeta}
    \xi=\overline{\gamma}\quad;\quad
    \zeta=\overline{\alpha\gamma}~.
\end{align}
Notice that complete description of non-Hermitian KEO's involves the third parameter $(\overline{\alpha}-\overline{\gamma})\neq0$. In table~\ref{TAB:xizeta} we report the values of linear parameters for KEO's presented in previous section. The equal values of $\xi$ and $\zeta$ for DA and BDD as well as for LK and W corroborate the findings of~\cite{dutra2000exact}.

\begin{table}[!ht]
\newcommand{\HideBeta}[1]{}
\begin{tabular}{lccccccl}\hline\hline
  KEO & $\vec{w}$ & $\vec{\alpha}$  \HideBeta{& $\vec{\beta}$} & $\vec{\gamma}$ &  $\xi$  & $\zeta$ & Ref.\\\hline
  vR  &$(\frac12,\frac12)$
        & $(\alpha,\gamma)$  \HideBeta{& $(\beta,\beta)$} & $(\gamma,\alpha)$ & $\frac12{(\alpha+\gamma)}$ & $\alpha\gamma$ &\cite{von1983position}\\
MB  &$1$
        & $\alpha$  \HideBeta{& $\beta$} & $\alpha$ & $\alpha$ & $\alpha^2$ &\cite{morrow1984model}\\
  BDD &$1$
        & $0$  \HideBeta{& $\!\!\!\!-1$}& $0$ & $0$ & $0$ &  \cite{bendaniel1966space} \\
ZK &$1$
        & $\!\!\!\!\!-\frac12$ \HideBeta{& $0$}& $\!\!\!\!\!-\frac12$ & $\!\!\!\!\!-\frac12$& $\frac14$ & \cite{zhu1983interface}  \\
  MM  &$1$
        & $\!\!\!\!\!-\frac14$ \HideBeta{& $\!\!\!\!\!-\frac12$}& $\!\!\!\!\!-\frac14$ & $\!\!\!\!\!-\frac14$& $\frac1{16}$ & \cite{mustafa2007ordering}\\
  GW &$(\frac12,\frac12)$
        & $(-1,0)$ \HideBeta{& $0$}& $(0,-1)$ & $\!\!\!\!-\frac12$ & $0$ & \cite{bastard1981superlattice}  \\
  LKDA\!\!\! &$(\frac12,\frac12)$
        & $(\alpha,0)$  \HideBeta{& $(\beta,\beta)$}& $(0,\alpha)$ & $\frac12\alpha$ & $0$ & \cite{dutra2000exact}\\
  LK &$(\frac12,\frac12)$
        & $(-\frac12,0)$  \HideBeta{& $(-\frac12,-\frac12)$}& $(0,-\frac12)$ & $\!\!\!\!-\frac14$ & $0$ & \cite{li1993band}\\
  W   &$ (\frac14,\frac12,\frac14)$
        & $(-1,0,0)$ \HideBeta{& $(0,-1,0)$}& $(0,0,-1)$ & $\!\!\!\!\!-\frac14$& $0$ & \cite{borges1988quantization}  \\
  DA  &$\frac{({\alpha},{\alpha},{1},1)}{2(\alpha +1)}$
        & $(-1,0,\alpha,0)$ \HideBeta{& $(0,0,\beta,\beta)$}& $(0,-1,0,\alpha)$ & $0$& $0$ & \cite{dutra2000exact} \\
  L\textit{al.}   &$ (\frac13,\frac13,\frac13)$
        & $(-1,0,0)$ \HideBeta{& $(0,-1,0)$}& $(0,0,-1)$ & $\!\!\!\!\!-\frac13$& $0$ & \cite{lima2012yet} 
       \\ \hline\hline
\end{tabular}
\caption{Linear parameters, $\xi$ and $\zeta$,  values for main KEO's encountered in the literatures. The weights $\vec{w}$  and the ordering ambiguity parameters $\vec{\alpha}$ , $\vec{\beta}$ and $\vec{\gamma}$ are as defined in~(\ref{TGEN}) along with the constraint~(\ref{Constraint}).}
\label{TAB:xizeta}
\end{table}


\section{Classification Scheme}

The two-parameter linear formulation we developed for Hermitian KEO's  agrees with dimensional arguments and analyticity conditions developed in~\cite{levy1995position}. However, we disagree with the conclusion therein that von Roos KEO is the most general Hermitian KEO.  This rather widespread~\cite{ganguly2006study,ganguly2006new,%
levy1995position,jaghoub2006effect,cavalcante1997form,%
dutra2003remarks,biswas2009coherent,%
kocc2003systematic,voon2002discontinuities,bagchi2006pseudo,%
mustafa2007ordering,galbraith1988envelope,dutra2000exact,%
bagchi2005deformed,gonul2002supersymmetric,chetouani1995green}, but incorrect, assertion in PDM literature  has its origin  in the fact that almost all of the so far proposed KEO's are (or equivalent to) special cases of von Roos form~(\ref{eq:TVR}).  This have led to inconsistent conclusions as in~\cite{ganguly2006new} when the authors managed to force a KEO  derived by Yan and Yee (YY) via path-integral formalism~\cite{yung1994derivation} to fit von Roos form~(\ref{eq:TVR}). The result was the intriguingly complex ordering ambiguity parameters with the inherent conceptual complication (non-Hermitian mass terms) as well as technical one (doubling the number of parameters). As we shall see, our formalism  and especially the explicit expression~(\ref{xizeta}) of the linear ordering ambiguity parameters in terms of von Roos ordering ambiguity parameters, provides a convenient and general framework to address such issues.

Making use of the Cauchy-Schwarz inequality together with the Hermiticity condition~(\ref{HermAG}) and the very reasonable  assumption~(\ref{0abg1}) yields
\begin{align}\label{xizetalim}
    \tfrac14\geq -\tfrac12\xi\geq\zeta\geq0~.
\end{align}
Inequalities~(\ref{xizetalim}) define the allowed region for Hermitian KEO's in the linear parameter-space $(\xi,\zeta)$. According to table~\ref{TAB:xizeta}, the linear ordering ambiguity parameters for von Roos KEO fulfill
\begin{align}\label{VRxizetaeq}
\xi^2\geq \zeta~.
\end{align}
For a given $T_{\xi,\zeta}$ fulfilling~(\ref{VRxizetaeq}) there is always  an equivalent von Roos KEO. However,  condition~(\ref{VRxizetaeq})  leaves a large part of the allowed region~(\ref{xizetalim}) incompatible with von Roos form. In other words, there is no equivalent von Roos KEO  for $\zeta> \xi^2$. This  is indeed the case for YY KEO for which $\xi=-\frac13$ and  $\zeta=\frac16$. So what is in this case the expression of those KEO's?

\begin{figure}[h]
\begin{center}
\includegraphics[width=15cm]{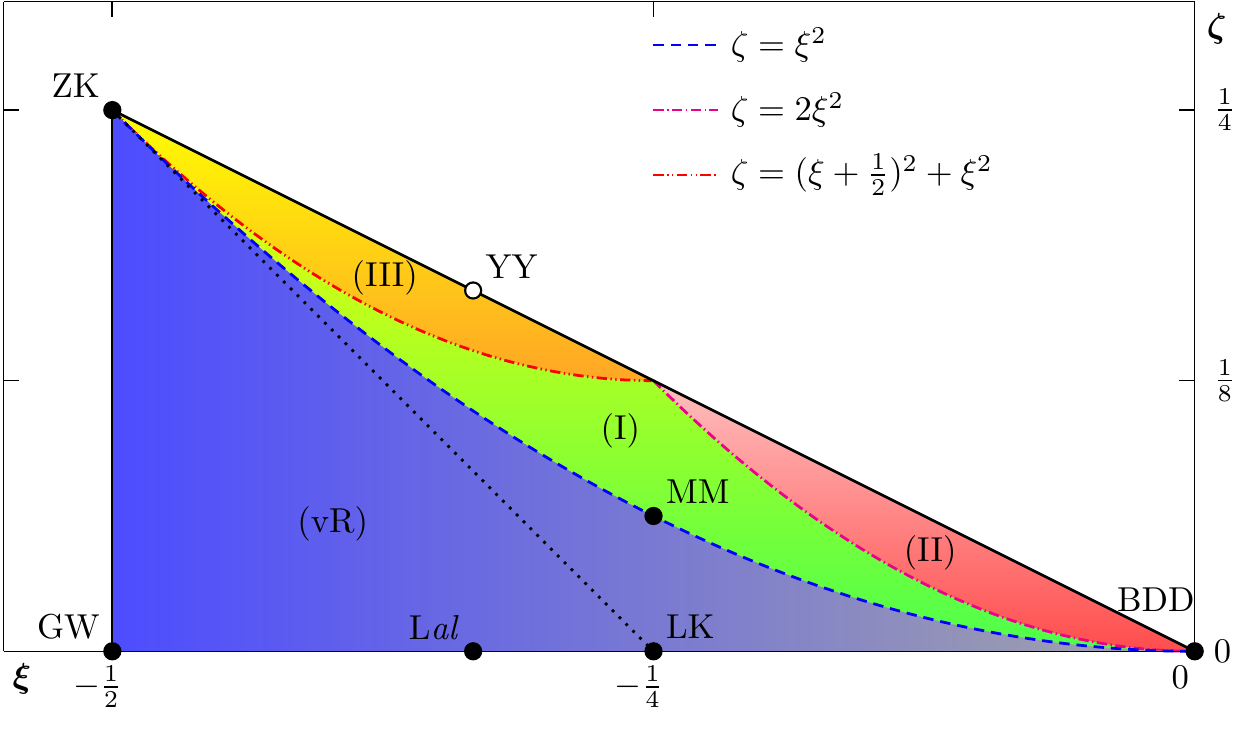}
\caption{\label{FIG:KEO_class}  KEO's types in the $(\xi,\zeta)$ plane. KEO's beyond the solid line limits do not comply with  $\alpha+\beta+\gamma=-1$ and/or $0\geq\alpha,\beta,\gamma\geq-1$. }
\end{center}
\end{figure}

Condition $\zeta\geq \xi^2$ defines in fact an entirely new family of manifestly Hermitian  KEO's of the form
\begin{align}\label{eq:TNew}
T_{\rm New}=\frac{1}{2}\!\!\left[w\: m^{\alpha}\vec{p}\,m^{\beta_1}\vec{p}\,m^{\alpha}+ (1-w)
m^{\gamma}\vec{p}\,m^{\beta_2}\vec{p}\,m^{\gamma} \right],
\end{align}
with  $2\alpha+\beta_1=-1$ and $2\gamma+\beta_2=-1$. Notice that $0\geq{\alpha}\geq-\frac12$ and $0\geq{\gamma}\geq-\frac12$ are totaly independent. As shown in figure~\ref{FIG:KEO_class}, this three-paramater family breaks down into three classes of two-parameter KEO's. The first one reads
\begin{align}\label{eq:TI}
T_{\rm I}=\frac{1}{4} \left[  m^{\alpha}\vec{p}\,m^{\beta_1}\vec{p}\,m^{\alpha}+
m^{\gamma}\vec{p}\,m^{\beta_2}\vec{p}\,m^{\gamma} \right]~,
\end{align}
for which, the linear ordering ambiguity parameters
satisfy
\begin{align}\label{NWxizetaeq}
\min\Big((\xi+\tfrac12)^2+\xi^2\:,\: 2\xi^2\Big)\geq\zeta\geq\xi^2~.
\end{align}
$T_{\rm I}$ have been first (de)considered by Morrow and Brownstein in~\cite{morrow1984model} without further investigation. The equality $\zeta= \xi^2$ holds for $T_{\rm MB}$ which appears to be a common  subclass for both  families.

The second new classes is defined within
\begin{align}\label{eq:RII}
-\frac12\xi\geq\zeta\geq2\xi^2
\end{align}
and has KEO's of the form
\begin{align}\label{eq:TII}
T_{\rm II}=\frac{1}{2} \left[ w\:m^{\alpha}\vec{p}\,m^{\beta}\vec{p}\,m^{\alpha} +(1-w)\:\vec{p}\,{\frac{1}{m}}\vec{p}
\right]
\end{align}
with $\tfrac12\geq w\geq0$. And finally the third class spans  the region
\begin{align}\label{eq:RIII}
 -\frac12\xi\geq\zeta\geq(\xi+\tfrac12)^2+\xi^2
\end{align}
with
\begin{align}\label{eq:TIII}
T_{\rm III}=\frac{1}{2} \left[w\:m^{\alpha}\vec{p}\,m^{\beta}\vec{p}\,m^{\alpha}+ (1-w)\:{\frac{1}{\sqrt{m}}}\,\vec{p}^2{\frac{1}{\sqrt{m}}}
\right]
\end{align}
and $\tfrac12\geq w\geq0$.

Figure~\ref{FIG:KEO_class} displays distinctively the different classes of KEO's  in the $(\xi,\zeta)$ plane.  For completeness, special cases are also reported. Table~\ref{TAB:xizetaN} summarizes the linear ambiguity parameters for the new family and its subsequent classes.

\begin{table}[!h]
\newcommand{\HideBeta}[1]{}
\begin{tabular}{lcccccc}\hline\hline
  Name & $\vec{w}$ \HideBeta{&$\vec{\alpha}$& $\vec{\beta}$} & $\vec{\alpha}$,$\vec{\gamma}$ &  $\xi$  & $\zeta$  \\\hline
  New &$(w,1\!-\!w)$
        \HideBeta{& $(\alpha,\gamma)$& $(\beta,\eta)$}& $(\alpha,\gamma)$ & ${w(\alpha\!-\!\gamma)+\gamma}$ & ${w(\alpha^2\!-\!\gamma^2)+\gamma^2}$        \\   I &$(\frac12,\frac12)$
        \HideBeta{& $(\alpha,\gamma)$& $(\beta,\eta)$}& $(\alpha,\gamma)$ & $\frac{\alpha+\gamma}{2}$ & $\frac{\alpha^2+\gamma^2}{2}$        \\   II &$(w,1\!-\!w)$
        \HideBeta{& $(0,\alpha)$& $(\beta,\beta)$}& $(\alpha,0)$ & $w\:{\alpha}$ & $w\:{\alpha^2}$        \\   III &$(w,1\!-\!w)$
        \HideBeta{& $(-\frac12,\alpha)$& $(\beta,\beta)$}& $(\alpha,-\frac12)$ & $w({\alpha}+\frac12)\!-\!\frac12$ & $w({\alpha}^2\!-\!\frac14)\!+\!\frac14$        \\         \hline\hline\hline
\end{tabular}
\caption{Linear parameters, $\xi$ and $\zeta$,  values for the new family and its subsequent classes. The weights and the ordering ambiguity parameters $\vec{\alpha}$ and $\vec{\gamma}$ are as defined in~(\ref{TGEN}) along with the constraints~$0\geq  \alpha,\gamma\geq-\frac12$ and~$\frac12\geq  w\geq0$.}
\label{TAB:xizetaN}
\end{table}

Similarly to MB one-parameter subclasses which is a common limit to vR and class-I (dashed line on figure~\ref{FIG:KEO_class}), the two other common limit one-parameter subclasses are
\begin{align}\label{eq:zeta2xi}
T_{\rm I/II }=\frac{1}{4} \left[ m^{\alpha}\vec{p}\,m^{\beta}\vec{p}\,m^{\alpha}+
\vec{p}\,{\frac{1}{m}}\vec{p}
\right]~,
\end{align}
for class-I  and class-II  (dash-dotted line) and
\begin{align}\label{eq:zetaxi2xi}
T_{\rm I/III}=\frac{1}{4} \left[ m^{\alpha}\vec{p}\,m^{\beta}\vec{p}\,m^{\alpha}
+{\frac{1}{\sqrt{m}}}\,\vec{p}^2{\frac{1}{\sqrt{m}}}
\right]~,
\end{align}
for class-I and class-III  (dash-double-dotted line). The linear formulation we developed above is clearly well adapted to discriminate between various KEO's in a model independent way up to limit cases.

As a concrete  application, let us go back to the puzzling KEO derived by Yan and Yee in the light of our classification scheme. As shown in figure~\ref{FIG:KEO_class}, this KEO  is definitely not vR-type. Since $\xi$ and $\zeta$ satisfy the constraint~(\ref{eq:RIII}), it is rather class-III with the real valued ordering ambiguity parameters
\begin{align}\label{eq:TYY}
T_{\rm YY}=\frac{1}{2} \left[\frac13\:\vec{p}\,{\frac{1}{m}}\vec{p}+ \frac23\:{\frac{1}{\sqrt{m}}}\,\vec{p}^2{\frac{1}{\sqrt{m}}}
\right]~.
\end{align}
This is in fact the first example of non-von Roos type KEO reported in the literature.

More generally, one no longer has to match form~(\ref{TGEN}) to built a Hermitian KEO for PDM system. One may now use alternatively the form~(\ref{Txizeta}) by choosing a couple of values for $(\xi, \zeta)$ fulfilling condition~(\ref{xizetalim}). Depending on the fulfilled constraint, the corresponding real valued ordering ambiguity parameters and weights can be calculated as indicated in table~\ref{TAB:AGxizeta}.

\begin{table}[!h]
\newcommand{\HideBeta}[1]{}
\begin{tabular}{lccc}\hline\hline
   Constraint & Type & ${w}$  & $\vec{\alpha}$,$\vec{\gamma}$    \\\hline
   Eq. (\ref{VRxizetaeq}) & vR &
    $ \frac12 $
        & $\left(\xi\pm\sqrt{\xi^2-\zeta},\xi\mp\sqrt{\xi^2-\zeta}\right)$ \\     Eq. (\ref{NWxizetaeq}) & I &
    $ \frac12 $
        & $\left(\xi\pm\sqrt{\zeta-\xi^2},\xi\mp\sqrt{\zeta-\xi^2}\right)$ \\    Eq. (\ref{eq:RII}) & II &
    $\xi^2/\zeta$
        & $(\zeta/\xi,0)$ \\    Eq. (\ref{eq:RIII}) & III &
    $\dfrac{(\xi+\frac12)^2}{\xi+\zeta+\frac14}$
        & $\left(\dfrac{\xi+2\zeta}{2\xi+1},-\dfrac12\right)$ \\\hline\hline
\end{tabular}
\caption{Ordering ambiguity parameters and weights as function of $\xi$ and $\zeta$.}
\label{TAB:AGxizeta}
\end{table}

\section{Duality between $T_{\rm vR}$ and $T_{\rm I}$}

\begin{figure}[h]
\begin{center}
\includegraphics[width=15cm]{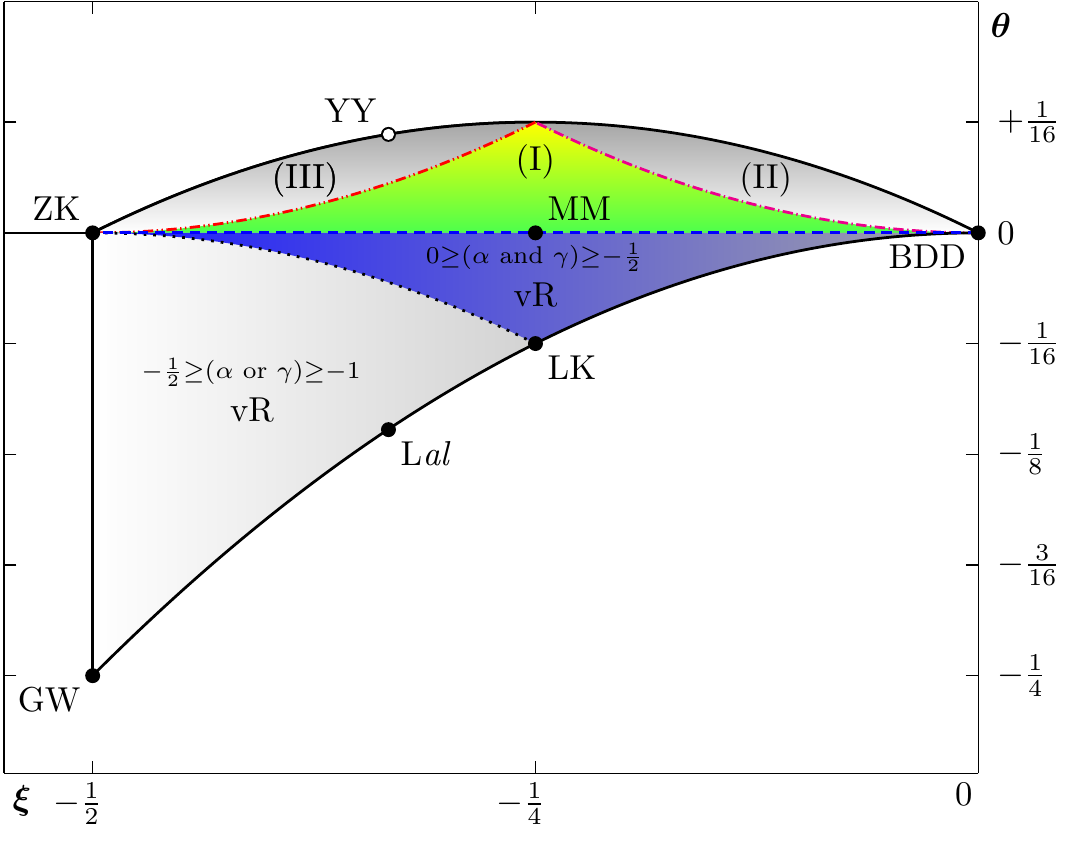}
\caption{\label{FIG:DUAL_class} Representation of the classes in the $(\xi,\theta)$ plane. The von Roos region dual to class-I was separated from the rest.}
\end{center}
\end{figure}

Let us now introduce the parameter
\begin{align}
 \theta=\zeta-\xi^2~.
\end{align}
Noting that $\theta_{\rm vR}\geq0$ and $\theta_{\rm I}\leq0$, we define the duality transformation $U$ by
\begin{align}
 U^{-1}T_{\theta,\xi}U=T_{-\theta,\xi}
\end{align}
The duality transformation $U$ transforms  the subset of vR-type KEO's for which  $0\geq\alpha,\gamma\geq-\frac12$   to class-I KEO's
\begin{align}
 U^{-1}T_{\rm vR}U=T_{\rm I}
\end{align}
and vice-versa, with exactly the same values of ordering ambiguity parameters $\{\alpha,\gamma\}_{\rm vR}=\{\alpha,\gamma\}_{\rm I}$. This duality is shown on figure~\ref{FIG:DUAL_class}. Note that Morrow and Brownstein KEO has the remarkable feature to be self-dual.

For $\psi_{\theta,\xi}$ solution and $E_{\theta,\xi}$ eigenenergy of  the Schrodinger equation
\begin{align}
  (T_{\theta,\xi}+V)\psi_{\theta,\xi}=E_{\theta,\xi}\;\psi_{\theta,\xi}
\end{align}
one would expect the solution and eigenenergy of the dual KEO to be
\begin{align}
  (T_{-\theta,\xi}+V)\psi_{-\theta,\xi}=E_{-\theta,\xi}\;\psi_{-\theta,\xi}
\end{align}
provided that boundary conditions and normalization could be properly handled. Since almost all the KEO's studied in the literature  are from von Roos family,  this duality would allow the extension of a number of the already published results to the new family.

\section{Conclusion}

In conclusion, we have introduced a generalized formulation and proved it was equivalent to two-parameter linear formulation in the case of Hermitician KEO's.  We proved that von Roos ordering is not the most general and built an entire new family of Hermitian KEO's.  This appears to be the first time that a complete classification of KEO's with PDM into two distinct families is pointed out. We were able to construct all the Hermitian KEO's and classify them into two-parameter classes. We have then presented the first  complete classification scheme for of KEO's with  position-dependent mass. We solved the puzzling case of Yang and Yee KEO and found it was in fact the first example of non-von Roos type KEO reported in the literature. A duality between von Roos ordering and one of the new classes  was identified. This duality would allow the extension of many existing results to the new family.

\section*{Acknowledgement}

This work has been supported by King Faisal University (DSR project 130191). A. Trabelsi is very grateful to T. Sbeouelji for numerous useful discussions.

\end{document}